\documentclass[12pt,fleqn]{article}
\usepackage{amsfonts,amssymb,cite}
\usepackage{epsfig,graphicx,color,latexsym}

\def\onecol{\onecolumn \mathindent 2em}
\def\noi{\noindent}

\makeatletter
\renewcommand{\section}{\@startsection{section}{1}{0pt}%
        {-3.5ex plus -1ex minus -.2ex}{2.3ex plus .2ex}%
        {\large\bf\protect\raggedright}}

\renewcommand{\subsection}{\@startsection{subsection}{2}{0pt}%
        {-3ex plus -1ex minus -.2ex}{1.4ex plus .2ex}%
        {\normalsize\bf\protect\raggedright}}

\renewcommand{\thesubsubsection}%
        {\arabic{section}.\arabic{subsection}.\arabic{subsubsection}.}

\renewcommand{\@oddhead}{\raisebox{0pt}[\headheight][0pt]{%
   \vbox{\hbox to\textwidth{\rightmark \hfil \rm \thepage \strut}\hrule}}}
\renewcommand{\@evenhead}{\raisebox{0pt}[\headheight][0pt]{%
   \vbox{\hbox to\textwidth{\thepage \hfil \leftmark \strut}\hrule}}}

\makeatother

\def\nqq{\hspace*{-2em}}
\def\nhq{\hspace*{-0.5em}}

\def\cm{\hspace*{1cm}}

\def\ten#1{\mbox{$\times 10^{#1}$}}

\def\Jl#1#2{{\it #1\/} {\bf #2},\ }

\def\ApJ#1 {\Jl{Astroph. J.}{#1}}
\def\CQG#1 {\Jl{Class. Quantum Grav.}{#1}}
\def\DAN#1 {\Jl{Dokl. AN SSSR}{#1}}
\def\GC#1 {\Jl{Grav. \& Cosmol.}{#1}}
\def\GRG#1 {\Jl{Gen. Rel. Grav.}{#1}}
\def\JETF#1 {\Jl{Zh. Eksp. Teor. Fiz.}{#1}}
\def\JETP#1 {\Jl{Sov. Phys. JETP}{#1}}
\def\JHEP#1 {\Jl{JHEP}{#1}}
\def\JMP#1 {\Jl{J. Math. Phys.}{#1}}
\def\NPB#1 {\Jl{Nucl. Phys.}{B\ #1}}
\def\NP#1 {\Jl{Nucl. Phys.}{#1}}
\def\PLA#1 {\Jl{Phys. Lett.}{#1A}}
\def\PLB#1 {\Jl{Phys. Lett.}{#1B}}
\def\PRD#1 {\Jl{Phys. Rev.}{D\ #1}}
\def\PRL#1 {\Jl{Phys. Rev. Lett.}{#1}}

\newcommand{\eqsection}{\makeatletter
    \@addtoreset{equation}{section}
    \renewcommand{\theequation}{\arabic{section}.\arabic{equation}}
    \makeatother}
\def\al{&\nhq}
\def\lal{&&\nqq {}}
\def\eq{Eq.\,}
\def\eqs{Eqs.\,}
\def\beq{\begin{equation}}
\def\eeq{\end{equation}}
\def\bear{\begin{eqnarray}}
\def\bearr{\begin{eqnarray} \lal}
\def\ear{\end{eqnarray}}
\def\earn{\nonumber \end{eqnarray}}

\def\nnn{\nonumber\\ \lal }

\def\yy{\\[5pt] {}}

\def\eql{\al =\al}

\def\dst{\displaystyle}

\def\fracd#1#2{{\dst\frac{#1}{#2}}}

\def\Half{{\fracd{1}{2}}}

\def\e{{\,\rm e}}
\def\d{\partial}

\def\diag{\mathop{\rm diag}\nolimits}

\def\const{{\rm const}}

\newcommand{\aver}[1]{\langle \, #1 \, \rangle \mathstrut}

\def\mn{_{\mu\nu}}
\def\mN{_{\mu}^{\nu}}
\def\cA{{\cal A}}

\def\vac{{}_{\rm (vac)}}

\def\lpl{l_{\rm pl}}

\def\Lem{Lema\^\i tre}
\def\Sch{Schwarzschild}
\def\dens{\ {\rm g/cm^3}}

\def\sph{spherically symmetric}
\def\ssph{static, spherically symmetric}

\begin{document}
\onecol
\thispagestyle{empty}

\section*
   {\Large\bf
    Multi-horizon spherically symmetric spacetimes with several scales of
    vacuum energy}

\bigskip
\noi{\bf Kirill Bronnikov$^{a,b,1}$, Irina Dymnikova$^{c,d,2}$,
        and Evgeny~Galaktionov$^d$}

\bigskip\noi {\small
    {\it $^a$ VNIIMS, 46 Ozyornaya St., Moscow, Russia}

\medskip\noi
    {\it $^b$ Institute of Gravitation and Cosmology,
        PFUR, 6 Miklukho-Maklaya St., Moscow 117198, Russia}

\medskip\noi
    {\it $^c$ Dept. of Math.\& Comp. Sci., Univ.of Warmia
        \& Mazury, S{\l}oneczna 54, 10-710 Olsztyn, Poland}

\medskip\noi
    {\it $^d$ A.F. Ioffe Physico-Technical Institute,
        Polytekhnicheskaja 26, 194021 St.Petersburg, Russia}
          }

\footnotetext[1]
    {E-mail: kb20@yandex.ru}
\footnotetext[2]
    {E-mail: irina@uwm.edu.pl}

\begin{abstract}
  We present a family of spherically symmetric multi-horizon spacetimes
  with a vacuum dark fluid, associated with a time-dependent and spatially
  inhomogeneous cosmological term. The vacuum dark fluid is defined in a
  model-independent way by the symmetry of its stress-energy tensor, i.e.,
  its invariance under Lorentz boosts in a distinguished spatial direction
  ($p_r=-\rho$ for spherical symmetry), which makes the dark fluid
  essentially anisotropic and allows its density to evolve. The related
  cosmological models belong to the \Lem\ class of models with anisotropic
  fluids and describe a universe with several scales of vacuum energy
  related to phase transitions during its evolution. The typical behavior
  of solutions and the number of spacetime horizons are determined by the
  number of vacuum scales. We study in detail a model with three vacuum
  scales: GUT, QCD and that responsible for the present accelerated
  expansion. The model parameters are fixed by the observational data and
  by analyticity and causality conditions. We find that our Universe has
  three horizons. During the first inflation the Universe enters a T-region
  which makes the expansion irreversible. After the second phase transition
  at the QCD scale the Universe enters an R-region, where for a long time
  its geometry remains almost pseudo-Euclidean. After crossing the third
  horizon related to the present vacuum density, the Universe should
  enter the next T-region with inevitable expansion.
\end{abstract}

{PACS numbers: 04.70.Bw, 04.20.Dw, 04.20.Gz, 98.80.Hw}
\vspace{4mm}

\section{Introduction}

  Astronomical observations give a compelling evidence for the existence of
  a dark energy dominating our Universe at above 73 \% of its density and
  responsible for its accelerated expansion due to negative pressure, $p =
  w\rho$, $w < -1/3$ [1--9]
  with the best fit $w=-1$ [10--15]
  which corresponds to the Einstein cosmological term $\lambda g\mn $
  related to the de Sitter vacuum $T\mn =8\pi G\rho\vac g\mn $
  ($\lambda = 8\pi G\rho\vac $).

  The well-known cosmological constant problem has two aspects: (i) Quantum
  field theory predicts for $\rho\vac$ the Planck scale $\rho\vac =
  \rho_{\rm Pl} = 5\times{10^{93}} \dens$.  Confronting this with the
  observational value, $\rho\vac \simeq 3.6\times{10^{-30}} \dens =
  1.4\times{10^{-123}}\rho_{\rm Pl}$, creates the Fine-Tuning Problem. (ii)
  The inflationary paradigm needs a large value of $\rho\vac $ at the
  beginning of the Universe evolution, typically of the GUT scale
  $\rho\vac\simeq{\rho_{\rm GUT}}\simeq 5\times 10^{77} \dens$; the
  observations indicate its much smaller value while the Einstein equations
  require $\rho\vac = \const$.

  A typical solution which can be found in the literature, is to put
  $\rho\vac = 0$ for some reason and to introduce a dark energy of
  non-vacuum origin which mimics the cosmological constant $\lambda$ when
  necessary. A lot of theories and models have been developed to describe a
  dynamical dark energy (for a comprehensive review see
  \cite{copeland1,copeland}). The alternative to the cosmological constant
  provided by quintessence assumes the existence of a hypothetical component
  of matter content with $w_Q \neq -1$ \cite{quint}. Q-models based on a
  scalar field, rolling down its self-interaction potential \cite{quint1},
  were tested using different methods with WMAP--CMB data. This gave the
  constraint $w_Q \leq -0.7$, with the best fit $w_Q = -1$ \cite{corasaniti}
  which evidently corresponds to the cosmological constant $\lambda$.

  Quartessence models describe a transition from a dust-dominated stage to a
  late-time inflationary stage. The Chaplygin gas model, with the postulated
  equation of state $p = -A/\rho$, gives a flat FRW model interpolating
  between $p=0$ and a negative pressure at late times \cite{quart}. It can
  be obtained in the model of a superfluid Chaplygin gas with the potential
  $V(\phi^*\phi) = M\left(\phi^*\phi/\mu + \mu/\phi^*\phi\right)$ which
  gives $p = -4M^2/\rho$ \cite{bilic,popov}. In the holographic dark energy
  approach (see for a recent review \cite{zhang}) with an interaction
  between dark matter with $p=0$ and dark energy \cite{hol}, quartessence
  can be recovered as an isotropic perfect fluid \cite{quart1}, but the
  perfect fluid for holographic dark energy was found to be classically
  unstable \cite{hol1}. The generalized Chaplygin gas (GCG) model, $p =
  -A/{\rho}^{\alpha}$ \cite{quart2}, was introduced to overcome difficulties
  with satisfying the CMB constraints \cite{quart3}. The observational
  constraint on the parameter $\alpha$, $0 < \alpha < 0.2$, implies a
  little difference between the GCG and $\lambda = \const$ \cite{quart4}.

  Quintom cosmology describes the dark energy in the framework of the
  brane-world cosmology by introducing two scalar fields, one being a
  quintessence and the other a phantom (for a review see \cite{quintom}).
  The multiple-lambda cosmology \cite{no2007} describes the Universe as a
  kind of multiverse \cite{multi} filled with phantom energy; it may
  describe the evolution as a sequence of transitions between different
  inflationary stages. It is based on a perfect isotropic fluid with a
  time-dependent equation of state \cite{no2005}, introduced
  phenomenologically \cite{no2007} and describing phantom-non-phantom
  transitions \cite{no2006}. One more approach to creating different
  effective scales of vacuum energy density rests on curvature-nonlinear
  multidimensional gravity with at least two extra factor spaces whose scale
  factors behave as scalar fields in four-dimensional space-time
  \cite{kb-rub-10}.

  According to observational data, dark energy is well described by the
  inflationary equation of state, $p=-\rho$. The stress-energy tensor has
  the form
\beq
    T\mN=\rho\vac \delta^{\nu}_{\mu},\ \ \ \rho\vac =\const.    \label{vac}
\eeq
  It represents a de Sitter vacuum which generates the de Sitter geometry
  responsible for accelerated expansion independently of specific properties
  of particular models of $\rho\vac $.

  The Standard Models of cosmology and particle physics suggest a series of
  phase transitions that have occurred in the course of the expansion and
  cooling history of the Universe \cite{boyanovsky}. A connection between
  particle physics and cosmology predicts the first inflation related to a
  phase transition at the GUT (Grand Unification) scale.  The first
  inflation solves the key problems of the standard Big Bang model
  (\cite{Olive} and references therein) and has been confirmed by CMB
  observations \cite{boyanovsky}.  The Standard Model of particle physics
  predicts another phase transition at the QCD (Quantum Chromodynamics)
  scale of 100 to 200 MeV (\cite{boyanovsky} and references therein) which
  can be related to a second inflationary stage \cite{boeckel}: A
  quasi-stable QCD vacuum state can lead to a short period of inflation (7
  e-foldings) consequently diluting the net baryon to photon ratio to its
  presently observed value.  The second inflationary stage is considered in
  the model of thermal inflation \cite{lyth}, with a duration of about
  10-foldings. Arguments for a second inflation at the QCD stage also exist
  in an effective model of a QCD phase transition which displays a high
  degree of supercooling at a critical temperature of the order of 100 MeV,
  so that the Universe increases exponentially during the quark-hadron
  transition \cite{borghini}.

  The aim of this paper is to present and study a family of cosmological
  solutions to the Einstein equations describing a vacuum-dominated universe
  which several scales of vacuum energy related to phase transitions in the
  course of its evolution.

  The gauge non-invariance of quantum cosmology leads to a connection
  between the gauge and the quantum spectrum of a certain physical quantity
  which can be specified in the framework of the minisuperspace model. There
  exists such a particular gauge in which the cosmological constant
  $\Lambda$ is quantized. Transitions between quantum levels of the operator
  $\Lambda$ can be related to several stages in the Universe evolution with
  different values of vacuum density $\rho\vac$ \cite{df}.

  At the classical level, the key point is the algebraic structure of the
  source term in the Einstein equations. In a model-independent approach, a
  vacuum is defined by the symmetry of its stress-energy tensor
  \cite{Gliner,id92,id2000}, as suggested by the Petrov classification for
  stress-energy tensors. The Einstein cosmological term corresponds to the
  maximally symmetric de Sitter vacuum with $\rho\vac =\const$.\footnote
       {Quantum field theory in curved spacetime does not contain a unique
        specification for the quantum state of a system, and the symmetry of
    the vacuum expectation value of a stress-energy tensor does not
    always coincide with the symmetry of the background spacetime
    \cite{Anderson}. In the case of de Sitter space, the renormalized
    expectation value of $\aver {T\mn}$ for a scalar field with an
    arbitrary mass $m$ and curvature coupling $\xi$ is proved to have a
    fixed point attractor behavior at late times (\cite{Anderson} and
    references therein), approaching, depending on $m$ and $\xi$, either
    the Bunch-Davies de Sitter-invariant vacuum or, in the massless
    minimally coupled case ($m=\xi=0$), the de Sitter-invariant
    Allen-Folacci vacuum. The latter case is peculiar since the de
        Sitter-invariant two-point function is infrared-divergent, and the
    vacuum states free of this divergence are O(4)-invariant Fock vacua;
    the vacuum energy density in the O(4)-invariant case is not the same
    (larger) as in the de Sitter-invariant case \cite{Kirsten}.}
  The Petrov classification of stress-energy tensors implies the existence
  of vacua whose symmetry is reduced as compared with (\ref{vac}), which
  allows the vacuum energy to become time-dependent and spatially
  inhomogeneous \cite{id2000,id2002,id2003}. A relevant class of
  stress-energy tensors describes a vacuum dark fluid \cite{DG2007}
  specified by the inflationary equation of state $p_{\alpha}=-\rho$ in only
  one or two distinguished spatial directions, so that the vacuum dark fluid
  is intrinsically anisotropic.

  In the spherically symmetric case, a cosmological vacuum is specified by
  $T^t_t = T^r_r$ ($p_r=-\rho$) \cite{id92,id2000}.  The radial direction is
  distinguished by the cosmological expansion.  Regular solutions with
  source terms specified by $T^t_t = T^r_r$, necessarily have a de Sitter
  center \cite{id2002}. In the case of two vacuum scales, at the center and
  at infinity, the source terms evolve from $\Lambda g\mn$ to $\lambda g\mn$
  with $\lambda \ll\Lambda$ \cite{id2000,id2002}. The cosmological models
  belong to the \Lem\ class of models with anisotropic pressures. They are
  asymptotically de Sitter at the early-time and late-time stages
  \cite{bdd,bd}.

  In this paper we study, in a general setting, spherically symmetric
  spacetimes with several vacuum scales. The relevant \Lem\ models involve
  several de Sitter (inflationary) stages in the Universe evolution. We
  study in detail a cosmological model with three basic vacuum scales: the
  GUT and QCD scales, and that responsible for the presently observed
  accelerated expansion, $\rho_{\lambda}=(8\pi G)^{-1}\lambda$.  We
  introduce a phenomenological density profile describing vacuum decay at
  each stage by an exponential function, as is typical for a decay, and fix
  the decay rate by the conditions of analyticity and causality.  This
  approach allows us to reveal certain general features of our Universe
  including the number of its spacetime horizons.

  The paper is organized as follows: In Sec.\,2 we introduce spherically
  symmetric vacuum spacetimes. In Sec.\,3 we show how the number of vacuum
  cales determines the general features of a spacetime, including the number
  of horizons. In Sec.\,4 we describe a transition from the static reference
  frame to geodesic reference frames representing \Lem\ cosmologies. Sec.\,5
  presents a \Lem\ cosmological model with a vacuum dark fluid,
  $T_t^t=T_r^r$. Sec.\,6 describes a model with three vacuum scales,
  GUT, QCD and present-day vacuum density, with the parameters fixed by the
  observational data. In Sec.\,7 we summarize and discuss the results.

\section {Vacuum energy in general \sph\ spacetimes}

  A general model-independent approach based on the Petrov classification of
  stress-energy tensors (SETs) defines a vacuum by the symmetry properties
  of its SET \cite{Gliner,id92,id2000}. The Einstein cosmological term
  corresponds to the de Sitter vacuum with the SET
\beq
    T\mN=\rho\delta^{\nu}_{\mu}, \qquad  p=-\rho.    \label{desit}
\eeq

  The medium specified by (\ref{desit}) is interpreted as a vacuum due to the
  algebraic structure of its SET (\ref{desit}). It has an infinite set of
  comoving reference frames, so that an observer cannot in principle measure
  his/her velocity with respect to it \cite{Gliner}, which is an intrinsic
  property of a vacuum \cite{landau}. The Einstein equations imply
  $\nabla_\nu T\mN =0$, which leads to $\rho=\const$ for the de Sitter
  vacuum (\ref{desit}). The maximum symmetry of the vacuum SET (\ref{desit})
  can be reduced while keeping its vacuum identity \cite{id92}, and this
  inevitably leads (due to $\nabla_\nu T\mN =0$) to a dynamical vacuum
  energy \cite{id2000}. The vacuum SET with a reduced symmetry, such that
  only one or two of its spatial eigenvalues coincide with the temporal
  eigenvalue, represents a vacuum dark fluid with the equation of state
  $p_{\alpha}=-\rho$ in the distinguished direction(s) \cite{DG2007}.

  The general time-dependent \sph\ spacetime is described by the metric
\beq                                                         \label{ds0}
     ds^2 = \e^{2\gamma}dt^2 - \e^{2\alpha}dR^2 - r^2 d\Omega^2,
     \qquad
        d\Omega^2 = d\theta^2 + \sin^2 \theta d\phi^2,
\eeq
  where $\alpha$, $\gamma$ and $r$ are functions of $R$ and $t$. The
  Einstein equations with source terms whose algebraic structure is
  specified by \cite{id92}
\beq
       T^t_t=T^r_r \ \ \ \ (p_r=-\rho)                       \label{vac1}
\eeq
  admit a class of regular solutions with a de Sitter center, $T\mN=
  \Lambda\delta^{\nu}_{\mu}$ as $r=0$, where $\Lambda=8\pi G\rho_0$
  corresponds to a certain fundamental scale of symmetry breaking
  $\rho_0 = \rho\vac$ at $r=0$ \cite{id2002,id2003}.

  In a comoving reference frame, the vacuum dark fluid specified by
  (\ref{vac1}) is presented by the SET
  \beq                                                    \label{Tvac}
     T\mN = \diag (\rho, \ \rho,\ -p_\bot,\ -p_\bot),
\eeq
  where $p_r$ and $p_\bot$ are the radial and transversal pressures,
  respectively.

  It can be easily shown that under these general conditions we necessarily
  have $\rho = \rho(r)$ and hence also $p_\bot = p_\bot(r)$. To begin with,
  the tensor (\ref{Tvac}) is invariant under any coordinate transformations
  in the $(t, R)$ 2D subspace, which is just a definitive property of a
  vacuum. Moreover, if there is no material source of gravity other than
  (\ref{Tvac}), the system satisfies all conditions of the generalized
  Birkhoff theorem \cite{bk80, bm95}, whence it follows that there exists a
  coordinate frame $(t, R)$ in which the metric (\ref{ds0}) is
  $t$-independent, and consequently $\rho$ and $p_\bot$ are functions of $R$
  alone. Let us show, however, that it is unnecessary to assume that
  (\ref{Tvac}) is the only source of gravity: it is sufficient to require
  that it does not interact with other kinds of matter, and thus  the
  conservation law $\nabla_\nu T\mN=0$ holds.

  Indeed, in this case we have (dots and primes stand for $\d/\d t$ and
  $\d/\d R$, respectively)
\bear                                                       \label{cons0}
      \dot \rho + 2\frac{\dot r}{r} (\rho + p_\bot) =0,
\\                                                          \label{cons1}
      \rho' + 2\frac{r'}{r} (\rho + p_\bot) =0.
\ear
  If $r=r(R)$, hence $\dot r=0$, \eq (\ref{cons0}) immediately gives
  $\rho=\rho(R)$. If, on the contrary, $r$ is $t$-dependent, so that in the
  most general case we can suppose $\rho = \rho(R, r)$, then (\ref{cons0})
  and (\ref{cons1}) combined lead to $\d\rho/ \d R =0$, as was asserted, and
  from (\ref{cons0}) it then also follows
\beq
      p_\bot = p_\bot (r)=-\rho-\frac{r}{2}\frac{d\rho}{dr}.  \label{pbot}
\eeq

  From (\ref{vac1}) it follows $G^0_0 = G^1_1$ due to the Einstein equations.
  In the static reference frame, using the Schwarzschild coordinate $R=r$,
  this gives $\alpha'+\gamma'=0$ in (\ref{ds0}), whence, choosing the
  appropriate time scaling, we get $\alpha + \gamma = 0$, and the metric
  (\ref{ds0}) takes the form
\beq                                        \label{ds1}
    ds^2 = A(r) dt^2 - \frac{dr^2}{A(r)} - r^2 d\Omega^2; \ \ \ \
    ~A(r)=e^{2\gamma(r)}
\eeq
  where the metric function $A(r)$ is given by
\beq
    A(r) = 1 - \frac{2M(r)}{r},                 \label{A(r)}
\eeq
  and $M(r)$ is the mass function
\beq
        M(r) = 4\pi\int_0^r \rho(x) x^2 dx.             \label{mass1}
\eeq

  We adopt the weak energy condition, i.e., a non-negative density for any
  observer on a time-like curve, which is natural for cosmological models
  describing the evolution of our Universe. This condition requires $p_\bot
  +\rho\geq 0$ which leads, by (\ref{pbot}), to a monotonically decreasing
  density profile $\rho(r)$ \cite{id2002}.  This fact, together with the
  number of vacuum scales at which $p_\bot$ satisfies $p_\bot=-\rho$,
  determines the generic behavior of the metric function $A(r)$, and as a
  result the maximum number of spacetime horizons. The actual number of
  horizons is determined by the specific form of the profile $\rho(r)$. In
  Section 6 we will introduce a density profile appropriate for three
  vacuum scales and show how the number of horizons in our Universe follows
  from the observational constraints.

  \eqs (\ref{vac1}) and (\ref{pbot}) give an $r$-dependent equation of
  state. It is evident that an anisotropic fluid needs two different
  equation-of-state parameters $w_{\alpha}=p_{\alpha}/\rho$. In our case,
  $w_r = p_r/\rho = -1$ due to \eq (\ref{vac1}), and $w_\bot = p_\bot/\rho
  = -1 - (r/2)d(\ln \rho)/dr$ due to \eq (\ref{pbot}). The parameter
  $w_\bot$ satisfies $w_\bot\geq -1$ since $\rho(r)$ is a monotonically
  decreasing function. The parameter $w_\bot$ approaches $w_\bot=-1$ as
  $r\to 0$, as $r\to\infty$ where $\rho$ approaches the present-day vacuum
  density  $\rho_\lambda= \const$, and also at each intermediate
  inflationary stage with $\rho\vac = \const$.

  The mass function (\ref{mass1}) can be related to the Schwarzschild mass
  if we separate in (\ref{mass1}) the presently observed vacuum density
  $\rho_{\lambda}$ as a background density. It is possible because $\rho(r)$
  is monotonically decreasing function, and $\rho_{\lambda}$ is its minimum
  value. If we introduce $\rho(r) = \rho_{d} + \rho_{\lambda}$, where
  $\rho_d$ is a dynamical density decreasing smoothly from the value at the
  center $\rho_0=(8\pi G)^{-1}\Lambda$ to zero at infinity, then the mass
  function (\ref{mass1}) takes the form $M(r) = 4\pi\int_0^r \rho_d(x) x^2
  dx + r^3\lambda/6$ and contains, in the limit $r\to\infty$, the
  Schwarzschild mass $M=4\pi\int_0^{\infty}{\rho_d(r)r^2dr}$ (in the
  Schwarzschild geometry it is measured by the Kepler law in the Newton limit
  of the Schwarzschild solution at its asymptotically flat infinity). The
  mass function (\ref{mass1}) differs from the proper mass in curved
  spacetime, which is obtained by integration with the proper volume element
  $dV = \sqrt{^3g}d^3x$, where $^3g$ refers to the determinant of the spatial
  metric. The difference represents the binding energy \cite{wald}, called
  also the gravitational mass defect \cite{landau}. The dynamics of \Lem\
  class models is determined by the mass function (\ref{mass1}), therefore
  we do not consider here the proper mass. Let us note, however, that in the
  case of a cosmology which is asymptotically de Sitter at infinity, the
  notion of the total conserved (proper) mass of the Universe can be
  introduced, because any asymptotically de Sitter spacetime must have an
  asymptotic isometry generated by the Killing vector $\d/\d t$, and there
  exists the notion of a conserved total mass of the spacetime as computed
  at the future infinity \cite{mann}.

\section {The number of horizons and the number of scales}

  A typical behavior of the metric function $A(r)$ in (\ref{ds1}) is
  dictated by the dynamics of the transversal pressure $p_\bot$ in its
  source (\ref{Tvac}), which determines the maximum number and nature of its
  extrema, determining in turn the maximum number of horizons.

  According to the Einstein equations, the transversal pressure $p_\bot$ can
  be expressed in terms of the metric function $A(r)$ as
\beq
     8\pi G p_\bot = \Half A'' + \frac {A'}{r}.            \label{p-bot}
\eeq
  At an extremum of $A(r)$, $A'=0$, hence, if $p_\bot > 0$, this extremum
  is a minimum, and this minimum of $A$ is unique in the domain where
  $p_\bot> 0$ (otherwise there would be a maximum between two minima).
  Assuming that $p_\bot$ is normally positive (so that the strong energy
  condition holds) and becomes negative only at distinguished spatial
  domains related to particular stages with a de Sitter vacuum behavior, we
  fix the number of zeros of $p_\bot$ and restrict the maximum number of
  zeros of $A(r)$, i.e., of spacetime horizons. One vacuum scale is related
  to the de Sitter center where $p_\bot$ is negative and $A(r)$ has a
  maximum. If there is no other such scale, the transversal pressure changes
  its sign once, and $A(r)$ has one minimum at which $A(r)$ can be negative.
  In the asymptotically flat case $A(r)\to 1$ as $r\to\infty$, hence one
  zero of $p_\bot$ can result in two zeros of $A(r)$, and the spacetime can
  have, as a maximum, two horizons \cite{id2002}. If there is also a de
  Sitter asymptotic at infinity, this gives another domain where $p_\bot$ is
  negative. Then $A(r)$ has two maxima and can have one minimum in between
  where $p_\bot$ is positive. In this case $p_\bot$ changes its sign twice,
  and the single minimum of $A(r)$ leads to at most 3 horizons \cite{bdd}.

  Each intermediate vacuum scale produces two more zeros of $p_\bot$ and at
  most two horizons. Hence, for $n$ vacuum scales with negative pressure we
  find at most $2n$ horizons in asymptotically flat spaces and at most
  $2n-1$ horizons in asymptotically de Sitter spaces. Note that it is the
  maximum number of horizons, and it will be smaller if, under the same
  behavior of $p_\bot$, the metric function is $A(r) > 0$ at some of its
  minima or $A(r) <0$ at some of its maxima.

  The dynamics of the transversal pressure also determines a typical
  behavior of the equation-of-state parameter $w_\bot$. For example, in the
  case of three vacuum scales of interest here, $\rho_{\rm GUT}$ at the
  center, $\rho_{\rm QCD}$ and $\rho_{\lambda}$, appearing successively,
  there can exist two domains where $p_\bot$ is positive. The parameter
  $w_\bot = p_\bot/\rho$ takes the value $w_\bot=-1$ during each
  inflationary stage, and it can be positive during the transitions
  $\rho_{\rm GUT}\to\rho_{\rm QCD}$, and $\rho_{\rm QCD}\to\rho_{\lambda}$.

  The requirements of regularity at the center and the dominant energy
  condition do not restrict the total number of horizons. This can be
  seen from the following example.

  Let $\rho > 0$ and consider two extreme equations of state compatible with
  the dominant energy condition:  {\bf (a)} $p_\bot = -\rho$ and {\bf (b)}
  $p_\bot =\rho$. In the case {\bf (a)} the most general \ssph\ solution is
  \Sch-de Sitter; in the case {\bf (b)} the SET structure is the same as for
  a radial electromagnetic field, so that we arrive at the Reissner-Nordstr\"om
  metric. The metric function $A$ in these two cases is
\bear
  {\rm {\bf (a)}}\qquad A(r) \eql 1 - \frac{2m}{r} - H^2 r^2,     \label{SDS}
\\                                    \label{RN}
  {\rm {\bf (b)}}\qquad A(r) \eql 1 - \frac{2\mu}{r} + \frac{q^2}{r^2},
\ear
  with constant parameters $m,\ \mu,\ H,\ q$. (Note that here $q$ is not a
  charge, the notation is adopted for convenience).

  Now, suppose that in the space-time with a regular de Sitter center
  the equation of state {\bf (a)}, leading to the exact de Sitter metric,
  holds in a finite interval of $r$ until $A$ reaches zero (that is, for $r
  < h_1 = 1/H$).  Beyond this horizon let us take the equation of state
  {\bf (b)}, so that the metric function $A$ has the form (\ref{RN}). It
  matches to the solution in the previous interval of $r$ if the constants
  $\mu$ and $q$ are found from the continuity conditions for $A$ and $A'$.
  Since $A'(h_1) < 0$, this necessarily means that $h_1$ is the inner
  horizon of the metric (\ref{RN}). With growing $r$ it will eventually
  reach the outer horizon $r=h_2$ with $A' > 0$. At this point we again
  switch the equation of state to {\bf (a)}, so that the next interval will
  be described by $A(r)$ given in \eq (\ref{SDS}), with $m$ and $H$
  determined from the continuity of $A$ and $A'$. At the inevitable next
  horizon $h_3$ we will again join the Reissner-Nordstr\"om metric in the same
  way and so on.

  The process can be continued as long as one wishes. Its feasibility is
  guaranteed by the following facts verified by a direct inspection:

  (i) Given any $h > 0$ and $C < 0$, one can always find such $\mu$ and $q$
  that the function (\ref{RN}) satisfies the conditions $A(h)=0$
  and $A'(h) = C$. Thus a next Reissner-Nordstr\"om segment can always be
  joined at the points $h_1$, $h_3$, etc.

  (ii) Given any $h > 0$ and $C > 0$ such that $hC <1$, one can always
  find such $m$ and $H$ that the function (\ref{SDS}) satisfies the
  conditions $A(h)=0$ and $A'(h) = C$; moreover, the condition $hC < 1$
  always holds if $h$ is the greater of two zeros of the function
  (\ref{RN}) and $C = A'(h)$ for the same function. Thus a next \Sch-de
  Sitter segment can always be joined at the points $h_2$, $h_4$, etc.

  The process can be stopped at any stage. If the last equation of state
  is {\bf (a)}, we obtain an asymptotically de Sitter model with an odd number
  of horizons; on the contrary, {\bf (b)} leads to an asymptotically flat
  model with an even number of horizons.

  The density profile is continuous but contains fractures (jumps of the
  derivative $\rho'$). Where $p_\bot =-\rho$, we have $\rho=\const$,
  while where $p_\bot = \rho$, the function $\rho(r)$ behaves as $1/r^4$.
  The fractures can, however, be smoothed by
  arbitrarily small additions to $\rho(r)$ without changing the whole
  qualitative picture, which will then correspond to an entirely smooth
  density distribution.

  Each plateau in the density profile must, from a physical viewpoint,
  manifest an intermediate energy scale, eventually connected with some
  phase transition. We can anticipate that the existence of such scales
  can appreciably complicate the set of possible spacetime structures.

\section {Transition to \Lem\ reference frames}

  Consider the general static metric (\ref{ds1}) that solves the Einstein
  equations with the SET (\ref{Tvac}). A transition to the geodesic
  coordinates $(R,\tau)$, where $\tau$ is the proper time along a geodesic
  and the radial coordinate $R$ is the congruence parameter, different for
  different geodesics, can be described in a general form. A radial timelike
  geodesic in the metric (\ref{ds1}) satisfies the equations
\beq
     \biggl(\frac{dr}{d\tau}\biggr)^2 = E^2 - A(r),     \qquad
     \frac{dt}{d\tau} = \frac {E}{A(r)},                   \label{geo'}
\eeq
  where the constant $E$ is connected with the initial velocity of a particle
  moving along this particular geodesic at a given value of the congruence
  parameter $R$. In general,
  $E = E(R)$, i.e., it is different for different geodesics.

\eqs (\ref{geo'}) give two of the four components of the
transition matrix
  $\|\d (t, r)/\d (\tau, R)\|$, namely, $\dot{r}$ and $\dot{t}$
  (dots and primes stand for $\d/\d\tau$ and $\d/\d R$, respectively)
  since this partial differentiation occurs along the geodesics:
\beq                                                       \label{rt-dot}
      \dot{r} = \pm \sqrt{E^2(R) - A(r)}, \cm
      \dot{t} = E(R)/A(r).
\eeq
  A relation between the other two components, $t'$ and $r'$, can be found
  from the condition $g_{\tau R} =0$ when we substitute $dt = \dot t d\tau
  + t' dR$ and $dr = \dot r d\tau + r'dR$ into the metric (\ref{ds1}):
\beq                                                       \label{t'}
       t' = \frac{\sqrt{E^2(R)-A(r)}}{E(R)A(r)} r'.
\eeq
  It remains to determine $r'(R, \tau)$, which can be done by using the
  integrability condition $(\d_\tau\d_R - \d_R\d_\tau)r =0$. The latter
  takes the form of a linear first-order differential equation with respect
  to $r' = y(R, r)$:
\beq                                                       \label{dr'}
       \d_r y = -\frac{y \ \d_r A}{2(E^2 -A)} + \frac{EE'}{E^2 - A}.
\eeq
  Solving it, we obtain
\beq                                                        \label{r'}
       y = r'(R, \tau) = \sqrt{E^2 - A}\Biggl[f_0(R)
        + EE'\int \frac{dr}{\big(E^2 - A(r)\big)^{3/2}}\Biggr].
\eeq
  The other integrability condition $(\d_\tau\d_R - \d_R\d_\tau)t =0$ holds
  automatically if (\ref{dr'}) holds. The functions $t(R, \tau)$ and
  $r(R,\tau)$ can now be found by further integration of \eqs
  (\ref{rt-dot})--(\ref{r'}). The resulting metric can be written as follows:
\beq                                                        \label{ds2}
       ds^2 = d\tau^2 - \frac{r'(R,\tau)^2}{E^2(R)} dR^2
                            - r^2(R,\tau) d\Omega^2.
\eeq

  One can see how this procedure works using de Sitter space as an example.
  Its static form is (\ref{ds1}) with $A(r) = 1 - H^2 r^2$, $H= \const$.
  We will choose three different families of geodesics such that
\beq
    E(R) = \sqrt{1 - K R^2}, \cm K = 0, \pm 1,              \label{E-dS}
\eeq
  and show that their corresponding reference frames represent the three
  well-known forms of the de Sitter metric as isotropic cosmologies with
  different signs of spatial curvature (see, e.g., \cite{Hawk}). Indeed,
  integrating the first relation in (\ref{rt-dot}) as an equation for $\tau =
  \tau(r, R)$ and properly choosing the arbitrary function of $R$ that
  appears as an integration constant, we obtain the following expressions for
  $r$:
\bear                                                       \label{r-dS}
       r (R, \tau) = (R/H) \times
            \Big\{ \cosh (H\tau),\ \e^{H\tau},\ \sinh (H\tau) \Big\},
\ear
  where the expressions in the curly brackets are ordered according to
  $K = 1,\ 0,\ -1$. Substituting them into (\ref{ds2}), we obtain the metric
  in the form
\bearr                                                       \label{dS}
       ds^2 = d\tau^2 - a^2(\tau)
            \biggl(\frac{dR^2}{1 - KR^2} + R^2 d\Omega^2 \biggr),
\nnn
    a(\tau) =
       (1/H) \times \Big\{ \cosh (H\tau),\ \e^{H\tau},\ \sinh (H\tau) \Big\},
\ear
  as was intended. One can also verify that the expression (\ref{r'})
  with $E(R)$ given by (\ref{E-dS}) (provided the function $f_0(R)$ is chosen
  properly) coincides with the expression for $r'$ obtained directly from
  (\ref{r-dS}) in all three variants. So the transition has been completed.

\section{\Lem\ cosmology with vacuum dark fluid $T^t_t=T_r^r$}

  We have shown above that the  behavior of the static metric function
  $A(r)$ is dictated by the number of vacuum scales, and that the static
  spherically symmetric metric (\ref{ds1}) can always be transformed to the
  \Lem\ form. Hence the cosmological evolution in this case can be described
  by a model from the \Lem\ class satisfying the condition $p_r = -\rho$
  that specifies a vacuum dark fluid, with the appropriate choice of the
  density profile $\rho(r)$ modelling smoothed jumps between different
  values of $\rho\vac $, which will be discussed in the next section.

  A \Lem\ class model is described by the line element \cite{landau}
\beq
    ds^2  = d\tau^2 -e^{2\nu (R,\tau)} dR^2
                    - r^2 (R,\tau) d\Omega^2,        \label{ds2Lem}
\eeq
  The coordinates $R, \tau$ are the Lagrange (comoving) coordinates. The
  coordinate $\tau$ measures the proper time along the world lines of a
  fluid. The function $r(R, \tau)$ corresponds to the Euler coordinate which
  is called luminosity distance.

  For the metric (\ref{ds2Lem}), the Einstein equations
  with the SET (\ref{Tvac}) read \cite{landau}
\bear
    8\pi G p_{r} \eql \frac{1}{r^2 }\left(e^{-2\nu}r'^2 -2r\ddot{r}-
           \dot{r}^2 -1\right),                  \label{eins1}
\yy                                                       \label{eins2}
    8\pi G p_{\bot} \eql \frac{e^{-2\nu}}{r}
        (r'' - r'\nu') - \frac{\dot{r}\dot{\nu}}{r}
        -\ddot{\nu} - \dot{\nu}^2 - \frac{\ddot{r}}{r},
\yy
    8\pi G \rho \eql -\frac{e^{-2\nu}}{r^2}
    \left(2rr''+r'^2 -2rr'\nu'\right)+\frac{1}{r^2 }
    \left(2r\dot{r}\dot{\nu}+\dot{r}^2 +1\right),        \label{eins3}
\yy
    8\pi G T^r_t \eql \frac{2 e^{-2\nu}}{r}
            \left( \dot{r}' - r'\dot{\nu}\right) =0,     \label{eins4}
\ear
  where dots and primes stand for $\d/\d\tau$ and $\d/\d R$.

  The component $T_t^r$ of the SET vanishes in the comoving reference
  frame since there is no momentum in the radial direction, and \eq
  (\ref{eins4}) is integrated giving \cite{landau,tolman}
\beq
        e^{2\nu} = \frac{r'^2 }{1+f(R)},                    \label{enu}
\eeq
  where $f(R)$ is an arbitrary function. Putting (\ref{enu}) into
  (\ref{eins1}), we obtain the equation of motion
\beq
    {\dot r}^2+2r{\ddot r}+ 8 \pi G p_r r^2=f(R).       \label{eqmotion}
\eeq
  Taking into account that
  $p_r + \rho =0$, the first integration of (\ref{eqmotion}) gives
\beq
    {\dot r}^2 = \frac{2GM(r)}{r} + f(R) + \frac{F(R)}{r},  \label{1int}
\eeq
  where the mass function $M(r)$ is defined by
\beq
        M(r) = 4\pi\int_0^r \rho(x) x^2 dx.             \label{mass}
\eeq

  The arbitrary function $F(R)$ (an ``integration constant'' parametrized
  by $R$) should be chosen equal to zero for models regular at $r = 0$
  since $M(r)\to 0$ as $r\to 0$ where $\rho(r)\to \rho_0< \infty$.

  The second integration of \eq (\ref{eqmotion}) gives
\beq
    \tau-\tau_0(R)=\int \frac{dr}{\sqrt{2GM(r)/r + f(R)}}.   \label{tau-lem}
\eeq
  The new arbitrary function $\tau_{0}(R)$ due to this integration
  is called the bang-time function \cite{silk}. For example, in the case of
  the Tolman-Bondi model for dust ($p_r = p_{\perp}=0$), the evolution is
  described by $r(R,\tau) = [9GM(R)/2]^{1/3}[\tau-\tau_0(R)]^{2/3}$, where
  $\tau_0(R)$ is an arbitrary function of $R$ representing the Big Bang
  singularity surface at which $r(R,\tau)=0$ \cite{CS}.

  In the presently considered regular case, asymptotically de Sitter in the
  R-region near $r=0$, the evolution starts from the timelike regular
  surface $r(R, \tau)=r_b$. For $f(R)\geq 0$, the bang surface is $r(R,
  \tau) =0$, and the solution (\ref{tau-lem}) near this surface reduces to
\beq
    \tau-\tau_0(R)=\int \frac{dr}{\sqrt{r^2/r_0^2 + f(R)}}, \label{taudes}
\eeq
  where
\beq
    r_0 = \sqrt{\frac{3}{8\pi G\rho_0}}                \label{r0}
\eeq
  is the curvature radius at $r=0$, and $\rho_0$ is the density at $r=0$. In
  the case $f(R) < 0$, the bang surface is $r(R, \tau) = r_b$ where $r_b$
  satisfies $2GM(r)/r + f(R)=0$. For small values of $f(R)$ we can apply
  (\ref{taudes}) which gives $r = r_0\sqrt{-f(R)}\cosh{[(\tau -\tau_0(R))
  /r_0]}$.  For $f(R) > 0$ we get $r = r_0\sqrt{f(R)}\sinh [(\tau-\tau_0(R))
  /r_0]$, and $r=r_0\exp[(\tau-\tau_0(R))/r_0]$ for $f(R)=0$.

  Different points of the regular timelike bang surfaces start at different
  moments of the synchronous time $\tau$, so that the bangs are non-singular
  and non-simultaneous.

  For $f(R)=0$ (parabolic motion), \eq (\ref{taudes}) gives
  at small $r$ the expansion law
\beq
    r = r_0 e^{(\tau-\tau_{0}(R))/r_0}                \label{rflat}
\eeq
  and
\beq
    e^{2\nu} = \frac{r^2 }{r_0^2}
             \left[\frac{d\tau_{0}(R)}{dR}\right]^2,      \label{enudes}
\eeq
The metric takes the FRW form with the de Sitter scale factor
 \beq
    ds^2 = d\tau^2 -r^2_0 e^{2c\tau/r_0}
                \left(dq^2 + q^2 d\Omega^2 \right),       \label{desflat}
\eeq
  where the variable $q = e^{\tau_0(R)/r_0}$ is introduced to transform the
  metric to the FRW form. In accordance with (\ref{rflat}), it describes a
  non-singular non-simultaneous de Sitter bang from the surface $r(\tau
  -\tau_0(R) \to -\infty) = 0$ \cite{us2001}.

  The inflationary stage is followed by an anisotropic Kasner-like stage:
  One scale factor, corresponding to the transversal direction, is given by
  $r(R,\tau)$, and the other, corresponding to the radial direction, is
  proportional to $r'$ according to (\ref{enu}), its particular form
  depending on the density profile $\rho(r)$ and the choice of arbitrary
  functions of $R$. If we choose $f(R)=0$ and $\tau(R)=R$, the metric can be
  approximated by \cite{us2001,bdd}
\beq
    ds^2=d\tau^2-(\tau+R)^{-2/3}K(R)dR^2
                    - L (\tau+R)^{4/3} d\Omega^2,       \label{ds2kas}
\eeq
  where $K(R)$ is a smooth regular function and $L$ is a constant.

  A similar behavior can be found for a density profile which approximates
  phase transitions with several scales of vacuum energy (see below). At
  each transition, an inflationary stage is followed by an anisotropic
  Kasner-like stage.

  The generic behavior of the considered \Lem\ class solutions is related to
  the function $A(r)$ expressed by (\ref{A(r)}) in terms of the mass
  function (\ref{mass}). Given the density profile $\rho(r)$, its specific
  form entirely determines the detailed properties of $A(r)$ and, in
  particular, the number of horizons.

\section{\Lem\ cosmology with GUT and QCD phase transitions}

\subsection{Basic features}

  According to the conventional scenario, the first inflationary stage
  corresponding to the GUT phase transition occurred at the GUT scale
  $E_{\rm GUT}\sim 10^{15}$ GeV, the relevant density being $\rho_{\rm GUT}
  \simeq 2.3\times 10^{77} \dens$. It was followed by a decay of vacuum
  energy resulting ultimately in a radiation-dominated stage. The next phase
  transition which could drive the second inflation [38--40]
  which occurred at the QCD scale $E_{\rm QCD}\sim (100\div 200)$ MeV, at
  about $10^{-5}$ seconds after the Big Bang, when the Hubble radius, $d_H =
  c/H$, was about 10 km \cite{boyanovsky}. The density $\rho_{\rm QCD}$ is
  smaller by a factor of $(E_{\rm QCD}/E_{\rm GUT})^4$ than the GUT density
  $\rho_{GUT}=\rho_0$, i.e., of the order of the nuclear matter density. The
  last inflationary stage corresponds to the presently observed dark energy
  density $\rho_\lambda$ which is about 107 orders of magnitude smaller than
  the GUT density.

  This situation can be modelled by the density profile
\beq
    \rho = \rho_0 \biggl[1 - (1 -B_1)\exp (-r_1^n/r^n)
        - (B_1 - B_3) \exp (-r_3^n/r^n)\biggr],           \label{profile}
\eeq
  where $n,\ B_1,\ B_3,\ r_1,\ r_3$ are constants, for which we adopt:
\beq
     B_1 = \rho_{\rm QCD}/\rho_0 \approx 10^{-64},\qquad
     B_3 = \rho_\lambda/\rho_0 \approx 10^{-107}; \qquad
     r_0 < r_1 \ll r_3.                                        \label{supp}
\eeq
  The exponential function in (\ref{profile}) is chosen as a typical one for
  decay processes. The parameter $n$ characterizing the decay rate will
  be fixed below by the conditions of analyticity and causality.

  We choose $f(R) \equiv 0$ in \eqs (\ref{enu})--(\ref{tau-lem}) because in
  this case each 3-hypersurface $\tau = \const$ is flat, with zero curvature
  \cite{bondi}, which guarantees fulfilment of the spatial flatness
  condition $\Omega=1$ required by the observational data.

  Under the above choice, the model undergoes the following stages:
\begin{description}                                       \itemsep 0pt
\item[(a)]
    $r \ll r_1$: the first inflation, $\rho \approx \rho_0$; the mass
    function (\ref{mass}) is approximated by $M(r)=\frac{4\pi}{3}r^3$,
    and \eq (\ref{tau-lem}) yields, in agreement with (\ref{rflat}),
\beq
        \tau-\tau_0(R) \simeq r_0 \ln{\frac{r}{r_0}}.       \label{tau_a}
\eeq

\item[(b)]
    $r \sim r_1$: end of the first inflation since the second term in
    (\ref{profile}) becomes significant.

\item[(c)]
    $r_1 \ll r \ll r_3$, so that
\beq                                                    \label{rho_c}
       \rho \approx \rho_0 (B_1 + r_1^n/r^n).
\eeq
    This stage in turn splits into two periods. As long as $r$ is
    sufficiently small,
\beq                                                    \label{c1}
       r < r_2, \cm  r_2 = r_1 B_1^{-1/n} = 10^{64/n} r_1,
\eeq
    the second term in (\ref{rho_c}) is dominant, so that $\rho(r)$
    rapidly decreases. It is an intermediate period between the first and
    second inflation. At $r = r_2$ the two terms coincide, and at
    $r > r_2$ we have $\rho \approx B_1 \rho_0 = \rho_{\rm QCD} = \const$,
    which corresponds to the second inflation.

\item[(d)]
    $r \sim r_3$: end of the second inflation since the third term in
    (\ref{profile}) becomes significant.

\item[(e)]
    $r \gg r_3$: the density is
\beq                                                      \label{rho_e}
        \rho \approx \rho_0 (B_3 + B_1 r_3^n/r^n).
\eeq
    Similarly to stage (c), at some value of $r$, namely, at $r = r_4$
    defined by
\beq                                                      \label{r4}
        r_4 = r_3 (B_1/B_3)^{1/n} = 10^{43/n} r_3 = 10^{107/n} r_1
\eeq
the two terms in (\ref{rho_e}) are equal. At $r_3 < r <
r_4$,
    we have one more intermediate period where $\rho(r)$ rapidly decreases,
    while at $r > r_4$ it approaches a constant corresponding to the
    present-day dark energy density.
\end{description}

  The time elapsed between the first and the second inflation,
  $\tau_2 - \tau_1$ (we denote $\tau_i = \tau(r_i)$) is estimated by
  integrating between $r_1$ and $r_2$ in \eq (\ref{tau-lem}). To this end,
  we find the mass function in the same interval:
\[
       M(r) = 4\pi
           \biggl(\int_{0}^{r_1} + \int_{r_1}^{r}\biggr)\rho(r)r^2 dr.
            \label{m11}
\]
  In the first term we take $\rho \approx \rho_0$ while in the second one,
  in accord with (\ref{rho_c}), we approximate $\rho(r)$ by
  $\rho_0 (r_1/r)^n$. Hence,
\beq                            \label{M11}
       M(r) = \frac{4}{3}\pi \rho_0 r_1^3
      + \frac{4\pi \rho_0 r_1^n}{n-3} \biggl(\frac{1}{r_1^{n-3}}
            -\frac{1}{r^{n-3}}\biggr).
\eeq
  In the last term, almost in the whole interval of interest, $r\gg r_1$,
  therefore for our estimation purpose we can neglect the last term thus
  obtaining a constant value of $M$,
\beq                            \label{M112}
      M(r) \approx M_2 = \frac{4\pi n}{3(n-3)}\rho_0 r_1^3 = \const.
\eeq
  Substituting it into (\ref{tau-lem}), we obtain
\beq                            \label{tau112}
      \tau_2 - \tau_1 \approx\sqrt{\frac{4(n-3)}{9n}}{r_0}
      \biggl(\frac{r_2}{r_1}\biggr)^{3/2}
\eeq

  The first restriction on the parameter $n$ is evident: $n > 3$.
  The second constraint follows directly from the dominant energy condition
  which requires $p_\bot \leq \rho$ and guarantees that the speed of sound
  never exceeds the speed of light, thus maintaining causality in the course
  of evolution. A simple analysis of \eq (\ref{pbot}) for several vacuum
  scales shows that the difference $\rho-p_\bot$ as a function of $r$
  decreases at each transition starting from $2\rho$ with $\rho = \const$.
  Let us introduce the function $f_{\rm DEC}=\rho-p_\bot$ characterizing
  the dominant energy condition.  According to (\ref{pbot}), $f_{\rm DEC} =
  2\rho+r\rho'/2$. It should be a decreasing function since $\rho(r)$ is
  monotonically decreasing, and its derivative $\rho'$ is negative. During
  the first transition, this function should decrease from $2\rho_0 =
  2\rho_{GUT}$ to $2\rho_{QCD}$. For the density profile (\ref{profile}) we
  have
\beq                             \label{dec1}
    f(r) = f_{\rm DEC}{\rho_0}^{-1} =  2\left[1-(1-B_1)e^{-r_1^n/r^n}
    \left(1+\frac{n}{4}\left(\frac{r_1}{r}\right)^n\right)\right]
\eeq
  It should be non-negative and decreasing from $2$ to $2B_1$, where
  $B_1=\rho_{QCD}/\rho_{GUT}$ is given by (\ref{supp}). For $r\gg r_1$ the
  exponent in (\ref{dec1}) can be presented as a series in $(r_1/r)^n$ which
  gives
\beq   \label{dec2}
    f=2B_1+2(1-B_1)\left[\left(1-\frac{n}{4}\right)
    \left(\frac{r_1}{r}\right)^n+\left(\frac{n}{4}
    -\frac{1}{2}\right)\left(\frac{r_1}{r}\right)^{2n}+ {\cal O}
        \left(\frac{r_1}{r}\right)^{3n}\right]
\eeq
  The condition of non-negativity of this function is $n\leq 4$. The
  derivative is given by
\beq            \label{decderivative}
    f'=\frac{(1-B_1)}{2}\frac{n}{r}\left(\frac{r_1}{r}\right)^n
    e^{-r_1^n/r^n}\left[n-4 -n\left(\frac{r_1}{r}\right)^n\right]
\eeq
  The condition $n\leq 4$ guarantees a monotonic decrease of the function
  $\rho-p_\bot$ during the first transition. It is easy to show that this
  concerns also the second transition at which the function $f_{\rm DEC}(r)$
  decreases from $2\rho_{\rm QCD}$ to $2\rho_{\lambda}$.

  The condition $n\leq 4$ thus provides non-negativity of the function
  $f_{\rm DEC}=\rho-p_\bot$ during the cosmological evolution with the
  density profile (\ref{profile}). Therefore we fix $n=4$ as the only
  integer compatible with analyticity and causality.

  Let us note that at the end of both transitions, for $r\gg r_1$ and $r\gg
  r_3$, the density in (\ref{profile}) behaves like $\rho\propto r^{-4}$, in
  a way typical of radiation in FRW cosmology where $\rho a^4 = \const$ and
  also agrees with our qualitative analysis in the second part of Section 3.

  With $n=4$ we get from (\ref{M11})
\beq                            \label{M1-2}
       M(r) = \frac{4}{3}\pi \rho_0 r_1^3
            + 4\pi \rho_0 r_1^4 \biggl(\frac{1}{r_1}-\frac{1}{r}\biggr).
\eeq
  The value of $M_2$ in (\ref{M112}) is now
\beq                             \label{M2}
      M_2 = \frac{16}{3}\pi \rho_0 r_1^3.
\eeq
  Substituting it into (\ref{tau112}), we obtain
\beq                                                \label{tau1-2}
      \tau_2 - \tau_1 \approx \frac{r_0}{3}
      \biggl(\frac{r_2}{r_1}\biggr)^{3/2} \sim 10^{24} r_0.
\eeq
  Recalling that $r_0 \sim 10^8\, \lpl$, we find $\tau_2-\tau_1 \sim 10^{32}
  \, \lpl/c \sim 10^{-11}$ s.  It is of interest that this estimate does not
  depend on the particular choice of the free parameter $r_1$ or,
  equivalently, the number of e-foldings $N_e := \ln(r_1/r_0)$.

  Furthermore, if the second inflation contains 7 e-foldings \cite{boeckel},
  it means that $r_3 \sim 10^3 r_2$. It is then also easy to find $r_4$, the
  value of $r$ at which the DE density has reached its modern value, from
  (\ref{r4}): $r_4 \sim 10^{43/4} r_3$.
To estimate the duration of the second inflation $\tau_3 - \tau_2$
and
  the time $\tau_4$ of the onset of the latest $\lambda$-dominated stage, we
  should integrate in (\ref{tau-lem}) from $r_2$ to $r_3$ and then to $r_4$.
  Acting in the same manner as in finding $\tau_2$, we see that the main
  contribution to the mass function comes from the range $r< r_2$ and is
  given by (\ref{M2}), hence the duration of the second inflation is
\beq                                                 \label{tau2-3}
     \tau_3 - \tau_2 \approx \sqrt{\frac 1{12\pi G\rho_0}}
            \biggl(\frac{r_3}{r_1}\biggr)^{3/2} \approx 10^{-7}\ {\rm s}.
\eeq
  Lastly, for $\tau_4$ we obtain the same relation as (\ref{tau2-3}) but
  with $r_3$ replaced by $r_4$. It results in
\beq                                                 \label{tau3-4}
    \tau_4 - \tau_3 \sim 10^{10}\ {\rm s} \sim 1000\ {\rm years}.
\eeq
  Let us note that all these times are practically independent of the number
  of e-foldings $N_e$ during the first inflation. However, the duration of
  the later period up to the present epoch does depend on $N_e$. Namely,
  since at $r \gg r_4$ we have $\rho \approx B_3 \rho_0 = \const$,
  integration in (\ref{mass}) at large enough $r$ gives
  \beq \label{mass4}
  M(r) \approx (4\pi/3) B_3\rho_0 r^3,
  \eeq
  and integration in (\ref{tau-lem}) yields immediately
\beq                                                   \label{tau5}
    \tau-\tau_0(R) \approx 10^{18}\ {\rm s}\cdot \ln (r/r_*),
\eeq
  where $r_* \approx 10^9 r_4 \approx 10^{36} r_1$ is the value of $r$ at
  which the contribution of $r^3$ to the mass function begins to exceed $M_2$.

  The qualitative behavior of the vacuum density profile (\ref{profile}) is
  shown schematically in Fig.\,1.
\begin{figure}[htp]
\centering
\vspace*{-4mm}
\epsfig{file=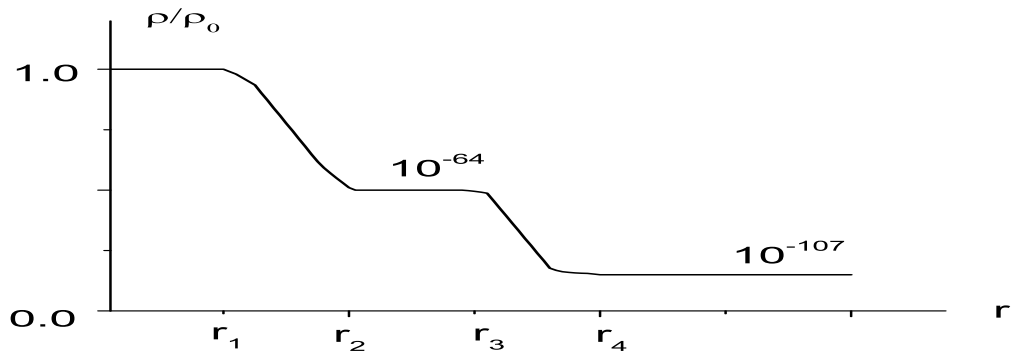,width=15.0cm,height=5.5cm}
\vspace*{-4mm}
\caption{\small Typical behavior of the vacuum density during the Universe
    evolution. Here $r_1$ is the end of the first inflation, $r_2$ and $r_3$
    are the beginning and end of the second inflation, respectively, $r_4$
    corresponds to achieving the present day vacuum density
    $\rho_{\lambda}$. } \label{fig.1}
\end{figure}

  \eq (\ref{tau5}) gives the expansion law $r = r_* e^{(\tau-\tau_{0}(R))
  /r_*}$, and \eq (\ref{enu}) gives for the second scale factor in
  (\ref{ds2Lem}) $e^{2\nu} = (r^2/r_*^2)(d\tau_0(R)/dR)^2$. Introducing the
  variable  $q = e^{\tau_0(R)/r_*}$, we transform the metric (\ref{ds2Lem})
  to the FRW form
\beq
        ds^2 = d\tau^2 -r^2_* e^{2c\tau/r_*}
                \left(dq^2 + q^2 d\Omega^2 \right),       \label{desflatNOW}
\eeq
  at the stage where the vacuum density achieves its present value.

  As we have seen, the only free parameter of the model is the number of
  e-foldings at first inflation $N_e = \ln (r_1/r_0)$, where the
  characteristic de Sitter radius for the GUT scale vacuum is $r_0 \simeq
  2.4\times 10^{-25}$ cm. The time interval corresponding to $r_1$ is,
  according to (\ref{tau_a}), approximately $10^{-35} N_e$ s. In the next
  subsection we evaluate an admissible interval for the parameter $N_e$ from
  the requirement of late-time homogeneity and isotropy.

  For the density profile (\ref{profile}) the transversal pressure is given by
\beq                            \label{pbotstory}
     p_\bot = \rho_0 \left[-1+(1-B_1)\biggl(1+\frac{2r_1^4}{r^4}\biggr)
     e^{-r_1^4/r^4} + (B_1-B_3)
     \biggl(1+\frac{2r_3^4}{r^4}\biggr)e^{-r_3^4/r^4} \right].
\eeq
  It satisfies the equation $p_\bot=-\rho$ during each inflationary stage.
  It has two maxima, $p_\bot\simeq 0.213\rho_{0}$ at $r_{m1}\simeq 1.2r_1$
  and $p_\bot\simeq 0.213B_1\rho_0=0.213\rho_{\rm QCD}$ at $r_{m2}\simeq
  1.2 r_2$ and ultimately quickly achieves $p_\bot=-B_3\rho_0=
  -\rho_{\lambda}$. The behavior of the transversal pressure
  (\ref{pbotstory}) is shown schematically in Fig.\,2.
\begin{figure}[htp]
\centering
\epsfig{file=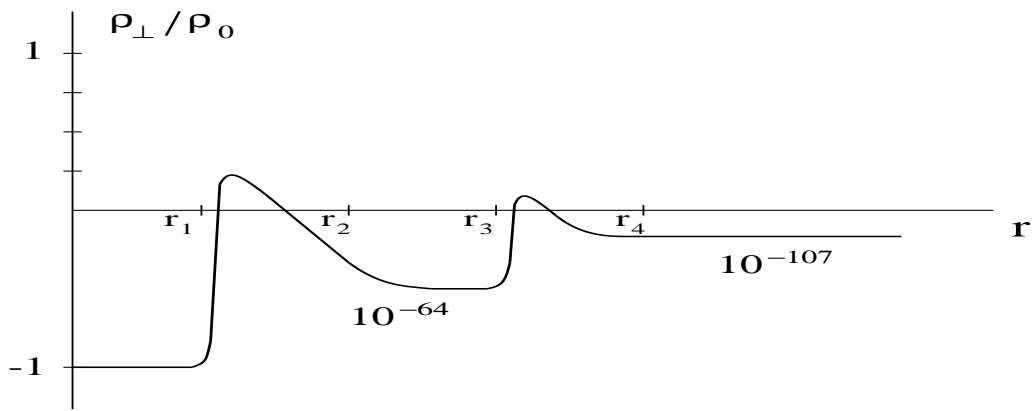,width=14.1cm,height=5.7cm}
\caption{\small Typical behavior of the transversal pressure during the
     Universe evolution. The quantities $r_1$--$r_4$ are the same as in
     Fig.\,1.} \label{fig.2}
\end{figure}

  The parameter $w_\bot = -1$ during the first inflation, then it rapidly
  increases to $w_\bot\simeq 0.213$ at $r_{m1}\simeq 1.2 r_1$, decreases to
  $w_\bot = -1$, quickly increases to $w_\bot\simeq 0.213$ at $r_{m2}\simeq
  1.2 r_3$, and finally approaches $w_\bot=-1$ as $\rho$ approaches
  $\rho_\lambda$.


\subsection{Late-time homogeneity and isotropy}

  At $r > r_4$ the model evolution is governed by the small effective
  cosmological constant $\lambda = B_3 \rho_0/(8\pi G)$ and tends to a de
  Sitter regime, i.e., becomes homogeneous and isotropic.

  The degree of inhomogeneity can be characterized by the dimensionless
  parameter $(r/\rho) d\rho/dr$ showing how the density $\rho$ changes at a
  distance $\sim r$. By (\ref{rho_e}) with $n=4$, at $r\gg r_4$ this
  parameter is approximately equal to $4r_4^4/r^4$ and rapidly decreases
  with growing $r$.

  Thus at $r \gg r_4$ one can estimate the degree of anisotropy of our model
  as is conventionally done for homogeneous models, e.g., using the
  anisotropy parameter \cite{harko02, dil-bi1-04}
\beq                                                        \label{cA-def}
      \cA = \frac{1}{3 H^2} \sum_{i=1}^{3} H_i^2 - 1,
\eeq
  where $H_i = {\dot a}_i/a_i$ are the directional Hubble parameters
  corresponding to the three scale factors $a_i(\tau)$, the dot stands for
  $d/d\tau$, and $H = (H_1+H_2+H_3)/3$ is the mean Hubble parameter (see
  \cite{dil-bi1-04} for a discussion of different anisotropy
  characteristics). In our model with the metric (\ref{ds2Lem}), where
  $e^{2\nu} = r'{}^2$ and $f(R)=0$, these scale factors are $a_1 = |r'|$
  and $a_2 = a_3 = r$. The expression for $r'$ is found from
  (\ref{tau-lem}):
\beq
    r' = - \sqrt{2G M(r)/r} \tau_0'(R).                  \label{r'anis}
\eeq
  Using this, one obtains for the anisotropy parameter
\beq                                                         \label{cA}
    \cA = 2 \frac{(\dot M/M - 3 \dot r/r)^2}{(\dot M/M + 3 \dot r/r)^2}
        = 2 \biggl[ \frac{3M_4 - 4\pi \rho_4 r_4^3}
                 {3M_4 + 4\pi \rho_4(2r^3 - r_4^3)}\biggr]^2,
\eeq
  where $M_4 = M(r_4)$ and $\rho_4 = \rho(r_4) = 2B_3 \rho_0$ according to
  (\ref{rho_e}). At large $r$ the parameter $\cA \sim r^{-6}$, but, as can
  be directly verified, this rapid decrease does not begin from $r_4$ but
  only from much larger values of $r$ because at $r\sim r_4$ both the
  numerator and the denominator of (\ref{cA}) are dominated by the constant
  $M_4$.

  To agree with CMB observations, the vacuum contribution must be already
  highly isotropic ($\cA < 10^{-6}$) when $r$ reaches the value of the scale
  factor $r = r_5 \sim 10^{25}$ cm corresponding to the recombination epoch
  with redshifts $z\sim 1000$. This requirement constrains the possible
  value of the free parameter $N_e = \ln(r_1/r_0)$. Indeed, the condition
  $\cA (r=r_5) < 10^{-6}$ gives
\beq                                                        \label{req-iso}
       3M_4 < 2\sqrt{2}\pi G\rho_4 r_5^3 \cdot 10^{-3}
\eeq
  (taking into account that $M_4 \gg \pi\rho_4 r_4^3$). In turn, $GM_4$ is
  expressed in terms of $r_1$. From (\ref{M2}) we know the value of $M_2 =
  M(r_2) = (16/3)\pi \rho_0 r_1^3$. To find $M_4$, we must integrate in
  (\ref{mass}) from $r_2$ to $r_4$; it turns out, however, that this
  integration contributes only a relative correction of the order $10^{-7}$
  to $M_2$. Thus
\beq                                                        \label{M4}
    M_4 \approx \frac{16}{3}\pi \rho_0 r_1^3.
\eeq
  Comparing (\ref{M4}) with (\ref{req-iso}), we obtain the constraint
\beq
      r_1 < B_3^{1/3}\,r_5/10 \approx \ten{-37} r_5
                = 3 \ten{-12}\ {\rm cm},
      \qquad     N_e = \ln (r_1/r_0) < 30.            \label{r1-iso}
\eeq

  Hence, after the recombination time corresponding to $r=r_5$, the \Lem\
  model (\ref{ds2Lem}) practically behaves as a homogeneous and isotropic
  FRW model (\ref{desflatNOW}).


\subsection{Evolution of the scale factors}

  Now, having established the constraint (\ref{r1-iso}), we can discuss
  the model evolution at all stages.

  For the spatially flat model, $f(R)=0$, the line element
  (\ref{ds2Lem}) takes the form
     \beq
    ds^2=d\tau^2-b^2(\tau, R)dR^2-r^2(\tau, R)d\Omega^2, \label{r,b}
\eeq
  where we have introduced explicitly two scale factors: $r(\tau, R)$ in
  accordance with (\ref{ds2Lem}) and $b(\tau, R)\equiv r'(\tau, R)$ in
  accordance with (\ref{enu}). For the integration ``constant'' in
  (\ref{taudes}) we choose $\tau_0(R)=-R$ to make the de Sitter asymptotics
  familiar. It is easily seen that in this case $\dot{r}(\tau +R) =
  dr/d(\tau+R)\d(\tau+R)/\d\tau = dr/d(\tau+R)$.

  Numerical integration of the \Lem\ equations during the first transition
  $\rho_{\rm GUT} \to \rho_{\rm QCD}$ shows an exponential growth of both
  scale factors at the beginning when $p_{\perp}\simeq p_r=-\rho$, followed
  by an anisotropic Kasner-like stage where the anisotropy of the pressures
  leads to an anisotropic expansion. The behaviors of the two scale factors
  $r(\tau+R), b(\tau+R)$ during the first phase transition are shown in
  Figs\,3 and 4.

\begin{figure}[htp]
\vspace{-1.0mm}
\centering
\vspace*{-4mm}
\epsfig{file=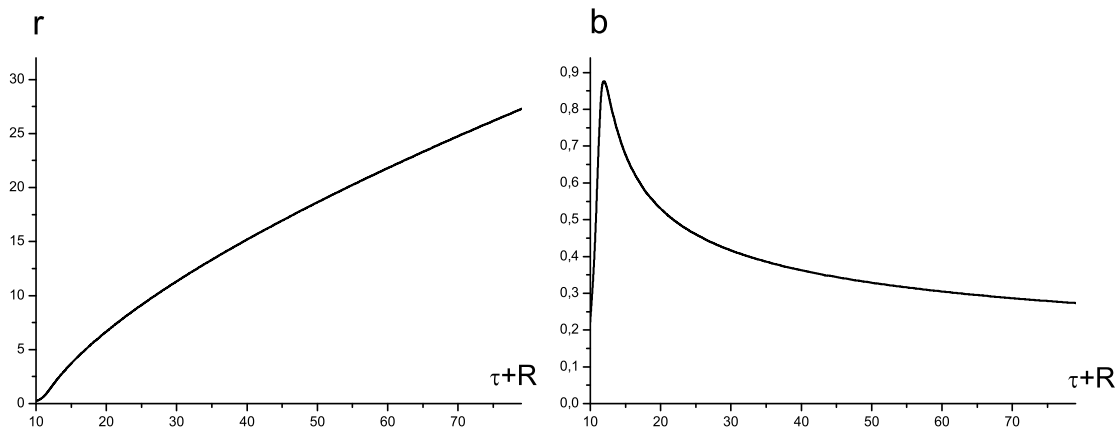,width=16.7cm,height=7.7cm}
\vspace*{-4mm}
\caption{\small The behavior of the scale factors during the first phase
     transition. The distances are normalized to $10^{12}r_0$ where
     $r_0\simeq {2.4\times 10^{-25}}$ cm is the characteristic GUT length
     scale for $M_{\rm GUT}\simeq 10^{15}$ GeV, and the time $\tau+R$ is
     normalized to the GUT time $t_{\rm GUT} = r_0/c\simeq 0.8\times
     10^{-35}$ s.} \label{fig.3}
\end{figure}
\begin{figure}[htp]
\vspace{-3mm}
\centering
\vspace*{-4mm}
\epsfig{file=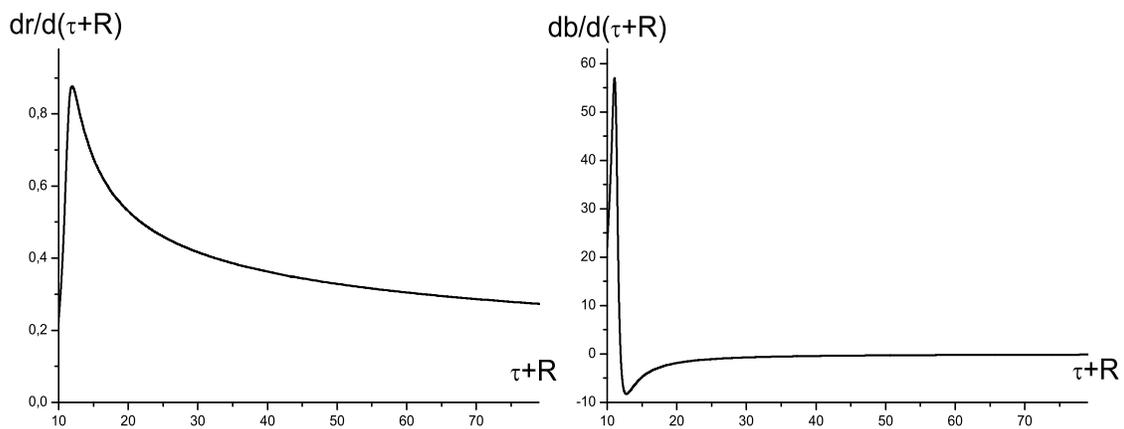,width=16.7cm,height=7.7cm}
\vspace*{-4mm}
\caption{\small The behavior of the velocities $\dot{r}(\tau + R)$,
      $\dot{b}(\tau+R)$ at the first transition.} \label{fig.4}
\end{figure}

  The behavior of the scale factors at the second phase transition is
  qualitatively quite the same for the density profile (\ref{profile}).

  The evolution of the scale factors during the whole Universe history is
  shown schematically in Fig.\,5 plotted on the basis of Fig.\,3 extended to
  the third inflationary stage (the hypersurface $\tau+R=t_4$ in Fig.\,5).
  The first inflation ends at the hypersurface $\tau+R=t_1$, the second
  inflation occurs between $\tau+R=t_2$ and $\tau+R=t_3$. The sharp maximum
  in ${\dot b}(\tau+R)$ in Fig.\,4, as well as the two maxima in the right
  panel of Fig.\,5 are related to a maximum of $p_\bot$ seen in Fig.\,2.  At
  late times, due to isotropy, the evolution of the two scale factors is
  common and conforms to the standard flat de Sitter cosmology.

\begin{figure}[htp]
\vspace{-3mm}
\centering
\epsfig{file=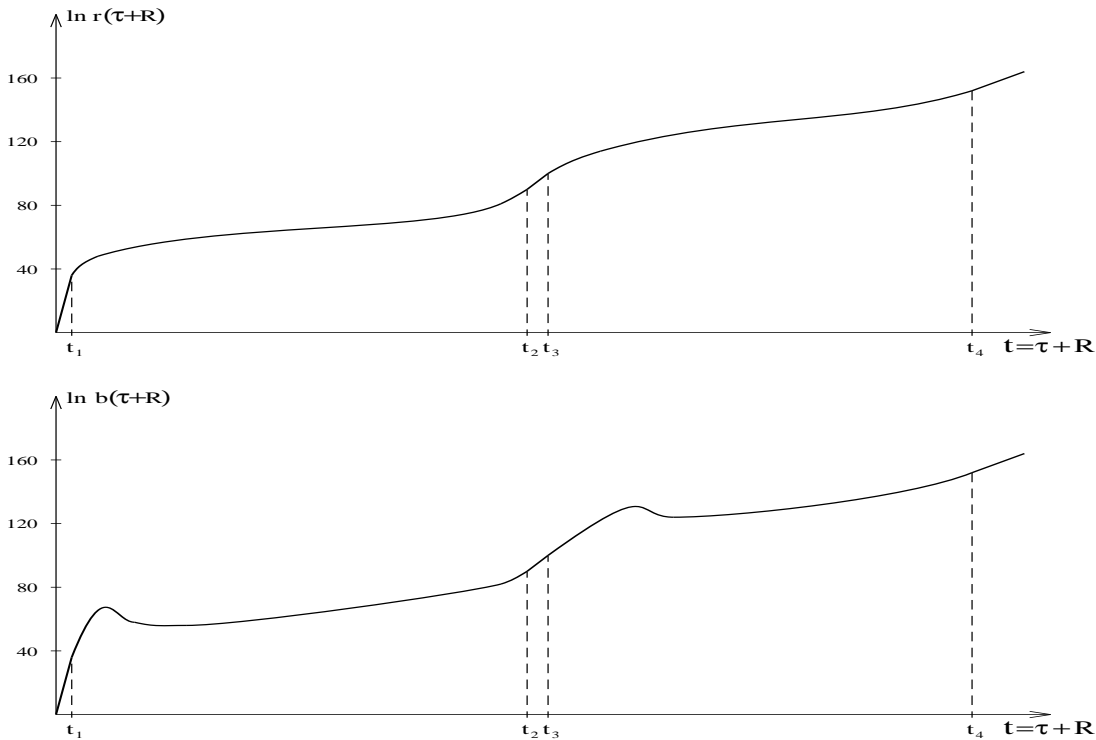,width=15.0cm,height=10.1cm}
\caption{\small Typical behavior of the scale factors $r(\tau+R)$ and
    $b(\tau+R)$.} \label{fig.5}
\end{figure}


\subsection{Horizons}

  We have described the Universe evolution from the viewpoint of the
  \Lem\ cosmological reference frame. Now we can find the number of horizons
  and present the global structure of spacetime. The proof in Sec.\,3
  concerned only the maximum number of horizons, and their smaller number is
  certainly possible, depending on the dispositions and durations of the
  phase transitions on the $r$ scale.

  Let us find how the whole scenario looks from the static reference frame.
  The metric function $A(r)$ calculated with the above fixed parameters  is
  shown in Fig.\,6, where the characteristic scales designated on the $r$
  axis are:

  $r_1 = 10^{-12}$ cm, $N_e = 29$ at the end of the first inflation.

  $r_2 = 1.2\times 10^4$ cm at the beginning of the second inflation,

  $r_3 = 1.2\times 10^7$ cm at the end of the second inflation,

  $r_4 = 6.7\times 10^{17}$ cm at achieving the present-day vacuum (dark
  energy) density  $\rho_{\lambda}$.

  The values of $r_5 = 10^{25}$ cm  and $r_6=10^{28}$ cm approximately
  correspond to the recombination and to the beginning of the third
  (presently observed) inflation, respectively.  The value $r_7 = 7.5\times
  10^{29}$ cm corresponds to the cosmological horizon due to de Sitter
  vacuum with the density $\rho_{\lambda}$.  The behavior of the metric
  function $A(r)$, Fig.\,6, testifies for the existence of three horizons
  for the case of the above fixed parameters corresponding to those of our
  Universe.
\begin{figure}[h]
\vspace{-4.0mm}
\centering
\epsfig{file=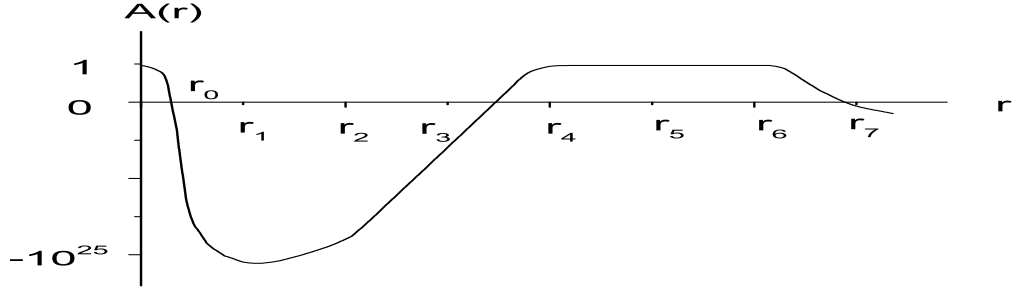,width=15.0cm,height=4.5cm}
\caption{\small The metric function $A(r)$ plotted with the parameters
    corresponding to our Universe} \label{fig.6}
\end{figure}

  The first horizon in our Universe, $r_{-}$, is close to the de Sitter
  radius $r_0$ corresponding to the GUT scale of the first phase transition.
  The second horizon $r_{+}\simeq 7.4\times 10^{13}$ cm distinguishes the
  additional essential length scale: for a long time our Universe evolves as
  the $T_{+}$-region $r_- < r < r_+$, i.e., the expansion is inevitable for
  all observers as dictated by the causal structure of spacetime. The
  present epoch gets into the the $R$-region between two $T_{+}$-regions.
  Near the point of achieving the present vacuum density the geometry
  generated by the vacuum fluid background becomes almost pseudo-Euclidean
  ($A(r)\simeq 1-10^{-6}$).

  The global structure of spherically symmetric spacetime with three
  horizons, asymptotically de Sitter in the center and at infinity,  is
  shown in Fig.\,7 \cite{bdd,castle1}.
\begin{figure}[htbp]
\centering
\includegraphics[width=10.7cm,height=8.1cm]{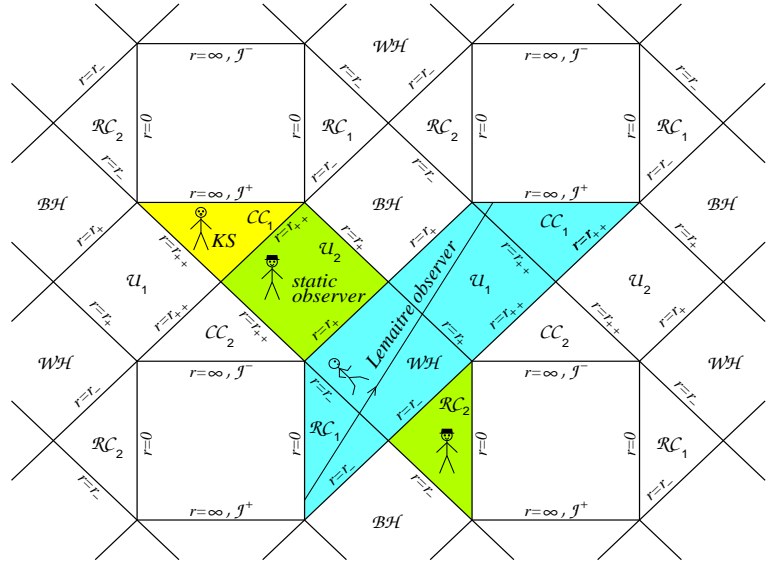}
\caption{\small Global structure of spherical space-time with three horizons}
    \label{fig.7}
\end{figure}
  This picture shows how the manifold of events is seen by different
  observers. Let us note that the Carter-Penrose diagram in Fig.\,7 covers
  the whole plane except for the squares bounded by the lines $r=0$ and
  $r=\infty$. The lines $r_{-}$ correspond to cosmological horizons for
  static observers (observers in hats) in the R-regions $0 \leq r < r_{-}$
  denoted as ${\cal RC}$; $r_+$ are the black (white) hole horizons for
  static observers in the R-regions $r_{+} < r < r_{++}$ denoted as ${\cal
  U}$, and $r_{++}$ are their cosmological horizons. T$_{+}$-regions $r_{++}
  < r < \infty$ denoted as ${\cal CC}$ correspond to regular homogeneous
  anisotropic cosmological T-models of Kantowski-Sachs type \cite{bdd,bd}.
  The \Lem\ cosmological model shown in Fig.\,7 starts its evolution in
  the R-region ${\cal RC}_1$ and goes consequently through the T$_+$-region
  $r_- < r < r_+$, the R-region $r_+ < r < r_{++}$ and the T$_+$-region
  $r_{++} < r < \infty$.

\section {Summary and discussion}

  We have presented a general analysis for the case of several scales of
  vacuum energy $\rho\vac$ corresponding to phase transitions involving
  inflationary stages. In our approach, the vacuum dark energy is described
  by a vacuum dark fluid defined by symmetry of its stress-energy tensor. In
  the spherically symmetric case, it is invariant under radial Lorentz
  boosts, acquiring the maximum symmetry of de Sitter vacuum only at
  inflationary stages. This makes the vacuum density $\rho\vac$
  time-dependent and spatially inhomogeneous. Cosmological solutions
  generated by the vacuum dark fluid belong to the \Lem\ class models with
  anisotropic pressures (their anisotropy follows directly from the
  variability of $\rho\vac$).

  The intrinsic properties of de Sitter space-time are responsible for an
  accelerated expansion, independently of particular properties of
  particular models of vacuum density associated with the cosmological
  constant. In a similar way, the intrinsic properties of geometries
  generated by a vacuum dark fluid can be in principle responsible for a
  variable vacuum density and make it possible to describe, on a common
  ground, the first inflationary expansion, the presently observed
  accelerated expansion, as well as inflationary stages related to phase
  transitions in the universe evolution, predicted by the Standard Model.

  To our knowledge, such an approach is applied for the first time for an
  analysis of the cosmological evolution in the case of several vacuum
  scales.

  The dynamics of cosmological models with a vacuum dark fluid is dictated
  by the number of vacuum scales: their number determines the behavior of
  the transversal pressure $p_\bot$, which in turn determines the maximum
  number of horizons; their actual number depends on the model parameters.

  We have studied in detail the cosmological model for the case of three
  vacuum scales: GUT, QCD and that responsible for the presently observed
  accelerated expansion. We used a phenomenological density profile with a
  typical behavior for a cosmological scenario with inflationary stages
  followed by a decay of vacuum energy, which we describe by an exponential
  function typical of decay processes. The parameter characterizing the
  decay rate is tightly fixed by the requirements of analyticity and
  causality. Other parameters of the model are fixed by the values
  $\rho_{\rm GUT}$, $\rho_{\rm QCD}$, $\rho_{\lambda}$ and $\Omega=1$. The
  only free parameter is the number of e-foldings in the first inflation,
  which is estimated using the observational constraint on the CMB anisotropy.

  This model reveals the following features of our Universe:

\medskip\noi
  (i) Our spacetime has three horizons.

\medskip\noi
  (ii) The cosmological evolution starts with a non-simultaneous timelike de
  Sitter bang followed by a short stage of inhomogeneous and anisotropic
  expansion.

\medskip\noi
  (iii) During the first inflationary stage, the Universe quickly enters
  a T$_+$-region which makes the expansion irreversible. The second phase
  transition occurs during this period.

\medskip\noi
  (iv) The Universe enters an $R$-region (in which we actually live) when
  the second inflation had already terminated. Soon after that the vacuum
  density reaches its present value (it occurs at $r \approx r_4 \sim
  10^{18}$ cm and $\tau \approx \tau_4 \sim 1000$ years). For a long time
  the Universe geometry remains almost pseudo-Euclidean up to the scale
  factor $r$ of approximately $3\times 10^{27}$ cm, which corresponds to an
  age of about $3\times 10^9$ years, when, according to the observational
  data, the present vacuum density $\rho_{\lambda}$ begins to dominate and
  the third inflation starts. Let us note that in a model taking into
  account matter and radiation, the vacuum density could achieve its present
  value later.

\medskip\noi
  (v) After crossing the third horizon related to the present vacuum
  density ($r_h \simeq 7.5\times 10^{29}$ cm), the Universe enters the
  second T$_+$-region with inevitable expansion.

\medskip
  We have seen that even our purely vacuum model can fairly well conform to
  the basic observational features of our Universe, which proves the
  viability of our approach. We hope that inclusion of matter and radiation
  in a further development of this approach will result in models more
  completely describing the observed cosmological picture.

  A general exact solution for homogeneous T-models has been found for a
  mixture of vacuum dark fluid with $T^t_t = T^r_r$ and dust-like matter
  \cite{bd}. It presents the class of T-models specified by the density
  profile of a vacuum fluid. The solution contains one arbitrary integration
  constant related to the dust density. Numerical estimates for a particular
  model illustrate the ability of such models to satisfy the observational
  constraints \cite{bd}. A similar solution for \Lem\ class models is
  now under consideration.

\subsection*{Acknowledgments}

  This work was supported by the Polish Ministry of Science and Education
  for the research project ``Globally regular configurations in General
  Relativity including classical and quantum cosmological models, black
  holes and particle-like structures (solitons)'' in the framework of the
  ``Polish-Russian Agreement for collaboration in the Field of Science and
  Technology''and by the Polish National Science Center through
  the grant 5828/B/H03/2011/40. KB acknowledges partial support from
  the grant NPK-MU (PFUR), RFBR grant 09-02-00677-a and by the Federal
  Purposeful Program ``Nauchnie i nauchno-pedagogicheskie kadry
  innovatsionnoy Rossii'' for the years 2009-2013. We are grateful to A.
  Dobosz for help with plotting Figs.\,2, 4 and 6.


\small
\begin{thebibliography}{99}  \itemsep 1pt

\bibitem{DE1}
    A. G. Riess et al 1998 {\it Astron. J.} {\bf 116} 1009

\bibitem{DE2}
    A. G. Riess et al 1999 {\it Astron. J.} {\bf 117} 707

\bibitem{DE3}
    S. Perlmutter et al 1999 {\it Astrophys. J.} {\bf 517} 565

\bibitem{DE4}
    N.A. Bahcall et al 1999 {\it Science} {\bf 284} 1481

\bibitem{DE5}
    L. Wang et al 2000 {\it Astrophys. J.} {\bf 530} 17

\bibitem{DE6}
    D.N. Spergel et al 2003  {\it Astrophys. J. Suppl. Ser.} {\bf 148} 175

\bibitem{DE7}
    E. Komatsu 2011 {\it Astrophys. J. Suppl.} {\bf 192} 18

\bibitem{DE8}
    M. Sullivan at al 2011 {\it Astrophys. J.} {\bf 737} 102; Arxiv:
        1104.1444

\bibitem{copeland1}
    E.J. Copeland, M. Sami, S. Tsujikawa 2006 {\it Int. J. Mod. Phys.}
        {\bf D 15} 1753

\bibitem{copeland}
    E.J. Copeland 2010, In {\it Proceedings of the Invisible
    Universe International Conference, Paris, France, 29 June--3 July
    2009} AIP: New York, NY, USA 132

\bibitem{lambda1}
    P.S. Corasaniti, E. Copeland 2002 {\it Phys. Rev. D}
        {\bf 65} 0430041

\bibitem{lambda2}
    S. Hannestad, E. Mortsell 2002  {\it Phys. Rev. D} {\bf 66} 0635081

\bibitem{lambda3}
    J.L. Tonry et al 2003  {\it Astrophys. J.} {\bf 594} 1

\bibitem{lambda4}
    J. Ellis 2003  {\it Phil. Trans. A} {\bf 361} 2607

\bibitem{lambda5}
    P.S. Corasaniti et al 2004 {\it Phys. Rev. D} {\bf 70} 0830061

\bibitem{lambda6}
    E.J. Copeland, M. Sami, S. Tsujikawa 2006 {\it Int. J. Mod. Phys.
     D} {\bf 15} 1753

\bibitem{quint}
    R.R. Caldwell et al 1998 {\it  Phys. Rev. Lett.} {\bf 80} 1582

\bibitem{quint1}
    I. Zlatev et al 1999  {\it Phys. Rev.} {\bf D 59} 123504

\bibitem{corasaniti}
    P.S. Corasaniti, E.J. Copeland 2002 {\it Phys. Rev.} {\bf D 65}
     043004; {\it ibid} 2004 {\bf D 70} 083006

\bibitem{quart}
    A.Y. Kamenshchik, U. Moschella, P. Pasquier 2001 {\it Phys. Lett.}
        {\bf B 511} 265

\bibitem{bilic}
    N. Bilic, G.B. Tupper and R.D. Viollier, Phys. Lett. {\bf B 535}
        (2002) 17

\bibitem{popov}
    V.A. Popov 2010 {\it Phys. Lett.} {\bf  B 686} 211

\bibitem{zhang}
    Z. Zhang, M. Li, X.-D.Li, S. Wang and W.-S. Zhang, arXiv:
        1202.5163 [astro-ph.CO]

\bibitem{hol}
    R. Horvat 2004 {\it Phys.Rev. } {\bf D 70 } 08730

\bibitem{quart1}
    W. Zimdahl 2007 {\it Int. J. Mod. Phys. }{\bf D 17}  651

\bibitem{hol1}
    Y.S. Myung 2007  {\it Phys. Lett.} {\bf B 652} 223

\bibitem{quart2}
    M.C. Bento, O. Bertolami and A.A. Sen 2002 {\it Phys. Rev.} {\bf D
        66}  043507

\bibitem{quart3}
    L. Amendola, F. Finelli, C. Burigana and D. Carturan 2003 {\it
        JCAP} {\bf 0307} 005

\bibitem{quart4}
    H. Hova and H.-X. Yang 2010 {\it USTC-ICTS-10-20, arXiv:1011.4788}

\bibitem{quintom}
    Yi-Fu Cai, E.N. Saridaks, M.R. Setara, J.-Q. Xia 2010  {\it Phys.
     Rep.}  {\bf 493} 1

\bibitem{no2007}
    S. Nojiri and S.D. Odintsov 2007 {\it Phys. Lett.} {\bf B 649} 440

\bibitem{multi}
    S. Robles-Perez, P. Martin-Moruno, A. Rozas-Fernandez and P.
       Gonzalez-Diaz 2007 {\it Class. Quant. Grav. } {\bf 24} F41

\bibitem{no2005}
    S. Nojiri and S.D. Odintsov 2005 {Phys. Rev.} {\bf D 72} 023003

\bibitem{no2006}
    S. Nojiri and S.D. Odintsov 2006 {\it Gen. Rel. Grav. Lett.} {\bf
     38} 1285

\bibitem{kb-rub-10}
        K.A. Bronnikov, S.G. Rubin, and I.V. Svadkovsky 2010 \PRD {81} 084010

\bibitem{boyanovsky}
    D. Boyanovsky, H.J. de Vega, D.J. Schwarz 2006 {\it Ann. Rev.
        Nucl. Part. Sci} {\bf 56} 441

\bibitem{Olive}
    K.A. Olive 1990 {\it Phys. Rep.} {\bf 190} 307

\bibitem{boeckel}
    T. Boeckel and J. Schaffner 2010 {\it Phys. Rev. Lett. }{\bf 105}
        041301; arXiv: 0906.4520

\bibitem{lyth}
    D.H. Lyth and E.D. Stewart 1995 {\it Phys. Rev. Lett.} {\bf 75}
        201; hep-ph/9502417

\bibitem{borghini}
    N. Borghini, W.N. Cottingham and R.V. Mau 2000 {\it J. Phys. G}
        {\bf 26} 771

\bibitem{df}
        I. Dymnikova and M. Fil'chenkov 2006 {\it Phys. Lett.} {\bf B 635}
    181

\bibitem{Gliner}
    E.B. Gliner 1965 {\it Sov. Phys. JETP} {\bf 22} 378

\bibitem{id92}
    I.G.  Dymnikova  1992 {\it Gen. Rel. Grav.} {\bf 24} 235

\bibitem{id2000}
        I.G. Dymnikova 2000 {\it Phys. Lett.} {\bf B 472} 33

\bibitem{Anderson}
    P.R. Anderson  et al. 2000 {\it Phys. Rev.} {\bf D 62} 124019

\bibitem{Kirsten}
    K. Kirsten and J. Garriga  1993 {\it Phys. Rev.} {\bf D 48} 567

\bibitem{id2002}
        I.G. Dymnikova 2002 {\it Class. Quantum Grav.} {\bf 19} 225

\bibitem{id2003}
    I. Dymnikova 2003 {\it Int. J. Mod. Phys. D} {\bf 12}  1015

\bibitem{DG2007}
    I. Dymnikova  and E. Galaktionov  2007 {\it Phys. Lett.}{\bf B
    645} 358

\bibitem{bdd}
        K.A. Bronnikov, A. Dobosz and I.G. Dymnikova 2003 {\it Class.
       Quantum Grav.} {\bf 20} 3797

\bibitem{bd}
        K.A. Bronnikov and I.G. Dymnikova 2007 {\it Class. Quantum Grav.}
     {\bf 24} 5803

\bibitem{landau}
    L.D. Landau and  E.M. Lifshitz 1975 {\it Classical Theory of
      Fields} Pergamon Press

\bibitem{bk80}
        K.A. Bronnikov and M.A. Kovalchuk 1980 {\it J. Phys. A: Math. Gen.}
        {\bf 13} 187

\bibitem{bm95}
        K.A. Bronnikov and V.N. Melnikov 1995 {\it Gen. Rel. Grav.} {\bf 27}
    465

\bibitem{wald}
    R.M. Wald 1984 {\it General Relativity}, Ch.6, Univ. Chicago Press

\bibitem{mann}
    A.M. Ghezelbashand R.B. Mann, {\it IHEP} {\bf 0201} (2002) 005

\bibitem{Hawk}
    S.W. Hawking, G.F.R Ellis 1973 {\it The large scale structure of
        space-time} Cambridge Univ. Press

\bibitem{tolman}
    R.C. Tolman 1934 {\it Proc. Nat. Acad. Sc. USA}  {\bf 20} 169

\bibitem{silk}
    D.W. Olson, and J.Silk 1979 {\it Ap. J.} {\bf 233} 395

\bibitem{CS}
    M.-N. Celerier, J. Schneider 1998 {\it Phys. Lett.} {\bf A249} 37

\bibitem{us2001}
    I. Dymnikova, A. Dobosz, M.L. Fil'chenkov,  A.A. Gromov 2001
        {\it Phys. Lett.} {\bf B 506} 351

\bibitem{bondi}
    H. Bondi 1947 {\it MNRAS}  {\bf 107} 410

\bibitem{harko02}
        T. Harko and M.K. Mak 2002 {\it Int. J. Mod. Phys.} {\bf D 11} 1171

\bibitem{dil-bi1-04}
        K.A. Bronnikov, E.N. Chudayeva, and G.N. Shikin 2004 {\it
        Class. Quantum Grav.} {\bf 21} 3389

\bibitem{castle1}
    I. Dymnikova 2004 in {\it Beyond the Desert 2003}, Ed. H.V.
     Klapdor-Kleinhaus; Springer Verlag:  Berlin, Germany; p 521;
        gr-qc/03100314.

\end{thebibliography}
\end{document}